\documentstyle[epsf,times]{mn2e}

\newcommand{\hub}[0]{km\,s$^{-1}$\,Mpc$^{-1}$}
\newcommand{\kms}[0]{km\,s$^{-1}$}

\newcommand{\pc}[1]{\protect\citename{#1}}
\newcommand{\dv}{$R^{1/4}\,$}

\newcommand{\mlu}{M$_\odot$/L$_{B,\odot}$}

\begin{document}

\title[The Hubble Constant from PG1115+080]{The internal structure of
the lens PG1115+080: breaking degeneracies in the value of the Hubble Constant}

\author[Treu \& Koopmans]{T. Treu$^1$ \& L.~V.~E.~Koopmans$^2$ \\ $^1$
California Institute of Technology, Astronomy, mailcode 105-24,
Pasadena CA 91125, USA \\ $^2$ California Institute of Technology, Theoretical Astrophysics,
mailcode 130-33, Pasadena CA 91125, USA}

\pubyear{2002}

\label{firstpage}

\maketitle

\begin{abstract}
We combine lensing, stellar kinematic and mass-to-light ratio
constraints to build a two-component (luminous plus dark) mass model
of the early-type lens galaxy in PG1115+080. We find a total mass
density profile steeper than $r^{-2}$, effectively $\rho\propto
r^{-\gamma'}$ with $\gamma'=2.35\pm0.1\pm0.05$ (random + systematic).
The stellar mass fraction is $f_*=0.67^{+0.20}_{-0.25}\pm0.03$ inside
the Einstein Radius (R$_{\rm E}$$\approx$1.2 effective radii). The
dynamical mass model breaks the degeneracies in the mass profile of
the lens galaxy and allows us to obtain a value of the Hubble Constant
that is no longer dominated by systematic errors:
H$_0=59^{+12}_{-7}\pm3$~\hub\ (68\% C.L.; $\Omega_{\rm m}=0.3$,
$\Omega_{\Lambda}$=0.7). The offset of PG1115+080 from the Fundamental
Plane might indicate deviations from homology of the mass profile of
some early-type galaxies.
\end{abstract}


\vspace{-0.3cm}
\section{Introduction}

Time delays between multiple images of gravitational lenses provide
the opportunity to measure the Hubble Constant H$_0$ (Refsdal 1964)
independently of local distance scale methods which rely on uncertain
calibrations (Saha et al.\ 2001; Freedman et al.\ 2001). Although the
method is attractive for its reliance on general relativity alone and
being a one-step global measurement, degeneracies inherent to the mass
distribution of the lenses seem hard to break without external
information (e.g. Koopmans 2001).

We recently showed that a combination of stellar kinematics and
gravitational lensing can be used to place tight constraints on the
mass distribution of early-type (E/S0) lens galaxies inside their
Einstein radius (Koopmans \& Treu 2002, Treu \& Koopmans 2002; KT02
and TK02). In particular, it was found that the total mass density
profiles of the lens galaxies in MG2016+112 and 0047--281 are very
close to $\rho\propto r^{-2}$. Both lens systems lie on the
Fundamental Plane (FP; Dressler et al.\ 1987; Djorgovski \& Davis
1987) at their respective redshifts, and are part of the sample that
we selected for the Lenses Structure and Dynamics (LSD) Survey,
currently being carried out at the Keck Telescope.

PG1115+080 ($z$=0.31; Weynman et al.\ 1980) is {\sl not} part of the
LSD Survey -- which focuses mainly on relatively isolated early-type
galaxies -- because of the presence of a massive compact group nearby
that could affect the mass distribution of the lens galaxy through
interaction. However, this system is particularly interesting because
time delays between the lensed quasar images ($z_{\rm s}=1.71$) are
available (Schechter et al.\ 1997; Barkana 1997), which can be used to
measure H$_0$. The geometry of this lens system has been modelled in
detail (Schechter et al.\ 1997; Keeton \& Kochanek 1997; Courbin et
al.\ 1997; Saha \& Williams 1997; Impey et al.\ 1998, hereafter I98;
Williams \& Saha 2000; Kochanek, Keeton \& McLeod 2001; Saha \&
Williams 2001; Zaho \& Pronk 2001), illustrating how strongly the
value of H$_0$ depends on the mass profile of the lens galaxy. Whereas
$\rho\propto r^{-2}$ mass models yield values of H$_0$$\la$50~\hub,
steeper or constant M/L mass models can yield values up to 60--70
\hub.

In this Letter, we apply the methods developed for the LSD Survey to
the lens E/S0 galaxy in PG1115+080 to break the mass-profile
degeneracy and perform an accurate measurement of the Hubble Constant,
minimising this dominant source of systematic uncertainty. Based on
the lensing and kinematic constraints (i.e.  the published velocity
dispersion measurement by Tonry 1998), we build a mass model (\S 2)
which we then use to determine H$_0$ (\S 3). We summarise and discuss
our results in \S 4.  Throughout, we define
$h={\rm H}_0/100$~km\,s$^{-1}$\,Mpc$^{-1}$, and we assume $\Omega_{\rm
m}=0.3$ and $\Omega_{\Lambda}=0.7$. The cosmography affects H$_0$ only
at a few percent level.

\vspace{-0.3cm}
\section{The Mass Distribution}

\label{sec:mass}

\subsection{Spectrophotometric properties of the lens}

\label{ssec:specphot}

Hubble Space Telescope (HST) images of the gravitational lens system
PG1115+080 are available from the HST archive. We select the images
with better resolution and sampling in each of the available bands:
$4\times640$s exposure with the Near Infrared Camera and Multi Object
Spectrograph (NICMOS; GO-7496; PI: Falco) Camera 2 (NIC2) through
filter F160W (I98); 10 exposures on the Planetary Camera (PC) of the
Wide Field and Planetary Camera 2 (WFPC2) through filter F814W for a
total exposure time of 15060 s (HST GO-6555; PI: Schechter);
$8\times400$s exposures on the WFPC2-PC through filter F555W (HST
G0-7495; PI: Falco)

The images were first reduced using a series of {\sc iraf} tasks based
on the {\sc drizzle} package (Fruchter \& Hook 2002).  This process
yielded maps of bad pixels, cosmic rays, saturated pixels, and pixels
affected by horizontal smearing for each individual exposure. Then, we
used a series of IDL scripts to subtract the QSO multiple images from
each individual exposure, after masking the deviant pixels. For each
exposure we used a library of subsampled synthetic Point Spread
Functions (PSFs) computed with TINY TIM 6.0 (Krist \& Hook 2001)
covering a range of the relevant parameters, such as jitter and focus
offset to simulate HST breathing. The PSF giving the smallest
residuals was used to remove the QSO images from each exposure.  The
individual exposures were finally combined to produce an image of the
lens galaxy.  The lens galaxy is imaged at high signal to noise ratio
in the F160W and F814W images, which we used to measure its structural
parameters. The structural parameters are listed in
Table~\ref{tab:photo}. An almost complete ring is visible in the F160W
image (see I98). Unfortunately, the signal-to-noise ratio is
significantly lower in the F555W exposure and structural parameters
could not be reliably measured. However, the colour
F555W--F814W=$1.82\pm0.02$ can be reliably measured within a fixed
aperture of radius $0\farcs6$.

Rest frame quantities were obtained as described in Treu et al.\
(2001b) using the F814W photometry, the F555W--F814W colour, and
correcting for galactic extinction with E(B--V)=0.041 (Schlegel,
Finkbeiner \& Davis 1998).  We use the parameters in the F814W filter
because it is much closer in wavelength than F160W to the standard
rest frame B and V bands, and the effects of possible colour gradients
are therefore minimised.

\begin{table}
\caption{Surface photometry of the lens galaxy. Rest frame quantities
through filters B and V are computed for $h=1$, $\Omega_m=0.3$,
$\Omega_{\Lambda}=0.7$ and E(B--V)=0.041 (Schlegel et al.\ 1998)}
\label{tab:photo}
\begin{tabular}{lrr}
  \hline
  & F160W & F814W \\
  \hline
  m (mag) & $16.36\pm 0.15$& $18.47\pm0.10$\\
  SB$_e$ (mag arcsec$^{-2}$) & $17.75\pm0.25$& $20.11\pm0.10$\\
  R$_e$  (arcsec) & $0\farcs76\pm0\farcs12$ & $0\farcs85\pm0\farcs07$\\
  \hline
  & V     & B  \\
  \hline
  M (mag) & $-20.89\pm0.12$ & $-20.05\pm0.12$ \\
  SB$_e$ mag arcsec$^{-2}$ & $19.83\pm0.12$ & $20.67\pm0.12$\\
  \hline
\end{tabular}
\end{table}

\begin{figure}
\begin{center}
\leavevmode
\hbox{%
\epsfxsize=1.05\hsize
\epsffile{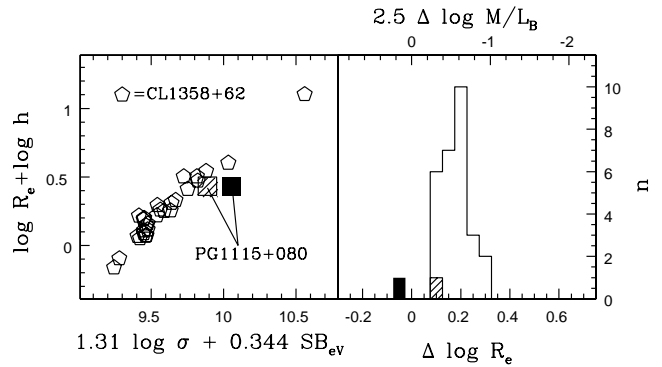}}
\end{center}
\caption{{\bf Left:} Comparison between the properties of the lensed
galaxy in PG1115+080 (filled square) and the FP of CL1358+62 (empty
pentagons). The hatched square is plotted using $\sigma_{\rm
SIE}=219\pm5$\kms as central velocity dispersion.  {\bf Right:} Offset
from the FP of Coma (zero), in the rest frame B band for the galaxies
in CL1358+62 and PG1115. Open histogram represents galaxies in
CL1358+62, solid and hatched histogram PG1115 as in the left panel.}
\label{fig:FP}
\end{figure}

Using the stellar velocity dispersion (Tonry 1998) of 281$\pm$25~\kms\
measured inside a $1\farcs0$ squared aperture, which corresponds to a
central velocity dispersion $\sigma=293\pm26$~\kms\ in a circular
aperture of radius $r_e/8$, we place the lens galaxy of PG1115+080 in
the FP space (i.e. the space with axis $\log \sigma$, SB$_e$, and
$\log {\rm R}_e$) and compare the position of the galaxy to the FP as
defined by the largest sample of galaxies at comparable redshift, 30
E+S0 galaxies in the cluster CL1358+62 at $z=0.33$ (Kelson et al.\
2000). As shown in the left panel of Fig.~\ref{fig:FP}, the lens
galaxy is dimmer than cluster galaxies with the same velocity
dispersion and effective radius (see also Kochanek et al.~2000).  The
lens galaxy is dimmer even than the galaxies in the Coma Cluster
(Bender et al.\ 1998) at variance with the observed trend that the
luminosity of E/S0 galaxies at a given radius and velocity dispersion
increases as a function of increasing redshift (e.g. Treu et al.\
2002). In the right panel of Fig~\ref{fig:FP} we show the residuals
from the FP of Coma for the lens galaxy and the galaxies in cluster
CL1358+62. The lens galaxy lies on the dimmest tail of the
distribution, and is offset by 4--5 times the rms from the
median/average CL1358+62 value. Since nothing in the morphology or
colours (e.g. the lens has B--V=$0.84\pm0.05$, whereas the average
colour of the galaxies in cluster CL1358 is B--V=0.86) of the lens
galaxy seems to indicate the presence of dust or anomalous stellar
populations, we will assume that the offset arises from structural
differences.

As in TK02 and KT02 we use the evolution of the effective mass-to-light 
ratio $\Delta \log (M/L_B)$, inferred from the local FP studies,
to estimate the stellar mass to light ratio $M_*/L_B$ of the lens
galaxy. Assuming that $\log (M_*/L_B)_{z}= \log (M_*/L_B)_{0} + \Delta
\log (M/L_B) \label{eq:FPev}$ (see discussion in Treu et al.\ 2001a;
Treu et al.\ 2002; KT02), we use the average local value
$12.0\pm4.1\,h$~\mlu\, (Gerhard et al.\ 2001 and references therein)
and the average evolution observed for field E/S0 galaxies $\Delta
\log M/L_B=-0.22\pm0.04$ at $z$=0.31 (Treu et al.\ 2002) to find
$M_*/L_B=7.2\pm2.5\,h$~\mlu\ for the lens galaxy.  The observed
evolution for cluster galaxies by van Dokkum et al.\ (1998) yields
$8.5\pm2.9\,h$~\mlu. In the context of our hypothesis of a structural
origin of the FP offset, we use the average observed values and {\sl
not} the evolution derived from the galaxy itself to estimate its
stellar mass to light ratio (which would yield
$M_*/L_B=14.6\pm5.1\,h$~\mlu).

\subsection{Dynamical and lensing models}

We calculate the expected stellar velocity dispersion of the lens
galaxy in PG1115+080 -- to compare then with the observed value, \S
2.1) -- by modelling its mass distribution with two spherical components
\footnote{See, e.g., Saglia, Bertin \& Stiavelli (1992 and Kronawitter
et al. 200) for discussions on the accuracy of the spherical
approximation.}, one for the luminous stellar matter and one for the
dark-matter halo. The luminous mass density profile is described
either by a Hernquist (1990; HQ) model or a by Jaffe (1983; JF) model.
The dark-matter distribution is modeled with a generalised version of
the Navarro, Frenk, \& White (1997; NFW) density profile. CDM
simulations show that the break radius ($r_b$) is much larger than the
Einstein Radius (e.g. Bullock et al. 2001). Hence, effectively
$\rho_d(r)\propto r^{-\gamma}$ inside the Einstein radius, where
$\gamma$ is the inner slope of the dark matter halo ($\gamma$=1 for
NFW). We further adopt an Osipkov--Merritt (OM; Osipkov 1979; Merritt
1985a,b) parametrisation of the anisotropy of the stellar velocity
ellipsoid, $\beta(r)=1-\sigma^2_{\theta}/\sigma_{r}^2
={r^2}/(r^2+r^2_i)$, where $\sigma_{\theta}$ and $\sigma_{r}$ are the
tangential and radial components of the stellar velocity dispersion
and $r_i$ indicates the anisotropy radius (see KT02 for
discussion). The line-of-sight velocity dispersion is determined by
solving the spherical Jeans equation, correcting for the average
seeing of 0\farcs8, and averaging the velocity dispersion profile --
weighted by the surface brightness -- inside the $1\,\sq''$
spectroscopic aperture.

A very similar mass model is used to reproduce the lensing geometry
with two minor differences: (i) the lens model allows for ellipticity
of the dark-matter halo ($b/a \ga 0.9$ in all cases) whereas the
luminous mass profile is spherical (consistent with the observations;
\S 2.1; I98); (ii) the luminous mass profile is modelled with a HQ or
pseudo-Jaffe profile (Keeton 2001). The latter is analytically
tractable and differs only marginally from the JF profile. An
additional component, accounting for the nearby massive compact group,
is necessary to reproduce accurately the lensing geometry
(e.g. Schechter et al 1997). We model the group as an isothermal
sphere (i.e. $\rho\propto r^{-2}$) or a NFW mass distribution.  The
break radius of the NFW profile is chosen to be 10$''$, consistent
with results from CDM simulations (e.g. Bullock et al.\ 2001) for the
observed velocity dispersion of the group ($\sim$330~\kms; Tonry
1998).

\subsection{Constraining the mass profile of the lens}

We now use the available observations to constrain the free parameters
of the models introduced in \S 2.2, with the overarching goal of
determining accurately the mass profile and hence the value of
H$_0$. To this aim we use four sets of observables. (i) The relative
astrometry of the multiple images and lens galaxy, derived from the
NICMOS images using PSF fitting, isophotal fitting, and centroiding
(Table~\ref{tab:astrom}; names are as in I98). The uncertainties
include the statistical component and the systematic difference
between the three techniques. Our astrometry is consistent within the
errors with that obtained by I98. (ii) The observations described in
\S 2.1, namely the effective radius, total luminosity, velocity
dispersion and stellar mass-to-light ratio of the lens galaxy. (iii)
The time delay between images B and C $\Delta t_{BC}=25.0\pm1.6$~days
(Barkana 1997; Schechter et al. 1997). (iv) The relative fluxes of the
images with 20\% errors (Table~\ref{tab:astrom}; see I98). The precise
choice of the errors on the flux-ratios are less important, because
the more precise image positions are the dominant contraints on the
lens mass model.
\begin{table}
\caption{NICMOS astrometry and relative fluxes. Plate scales are
$0\farcs0760261$/pix and
$0\farcs0753431$/pix along x and y respectively, taken from the
NICMOS history tool available at STScI web site. The position angle of
the NICMOS image is 68.7594 degrees.}
\label{tab:astrom}
\begin{tabular}{lccc}
  \hline
  Object & $\Delta$RA & $\Delta$ DEC & $S_{\rm H}^i/S_{\rm H}^{\rm C}$\\
  \hline
  A1  &   $+1\farcs328\pm0\farcs006$  & $-2\farcs042\pm0\farcs006$ & 2.29$\pm$0.08\\ 
  A2  &	  $+1\farcs472\pm0\farcs005$  & $-1\farcs581\pm0\farcs005$ & 3.44$\pm$0.15\\
  B   &	  $-0\farcs338\pm0\farcs006$  & $-1\farcs965\pm0\farcs005$ & 0.76$\pm$0.04\\
  C   &	  $+0\farcs000\pm0\farcs006$  & $+0\farcs000\pm0\farcs004$ & 1.00\\ 
  G   &	  $-0\farcs381\pm0\farcs007$  & $-1\farcs345\pm0\farcs007$ &  --\\
 \hline
\end{tabular}
\end{table}

Specifically, we use these observations and the dynamical model to
constrain the slope of dark matter halo ($\gamma$) and the fraction of
stellar mass within the Einstein Radius ($f_*$). The relative
astrometry yields the projected mass of the lens galaxy within the
Einstein Radius -- corrected for the projected mass contribution of
the group -- which is virtually independent of the mass profile of the
galaxy or that of the nearby group for the two group mass profiles
that we adopt (see \S 2.2). 

We find a mass of $M(R<{\rm R}_E=1\farcs04)=(1.19\pm 0.06) \times
10^{11}\,h^{-1}$~M$_\odot$ -- corresponding to $\sigma_{\rm
SIE}=219\pm5$~\kms\, for a singular isothermal ellipsoid -- consistent
with results from I98.  The Einstein radius R$_{\rm E}$=$1\farcs04$
corresponds to R$_{\rm E}= 3.32\,h^{-1}$\,kpc at the redshift of the
lens. The half-light radius of the luminous component is set equal to
the effective radius of the best fitting \dv model (\S 2.1). For any
given value of the anisotropy radius $r_i$, we determine the
likelihood of each ($f_*$,$\gamma$) pair, by comparing the observed
stellar velocity dispersion with that from the dynamical model, and
assuming Gaussian error distributions. In the limit of negligible
stellar mass fraction ($f_*\rightarrow 0$), the ``effective slope''
(see TK02 or KT02) of the lens-galaxy mass profile is $\gamma'=2.38\pm
0.08$ and $\gamma'=2.31\pm 0.08$ (68\% CL) for the isotropic
($r_i=\infty$) HQ and JF luminous profiles, respectively. These values
are lower by 1--2\% for non-isotropic models with $r_i=R_{\rm
e}$. This suggests that a steep total mass profile is required for the
lens galaxy to explain the large observed stellar velocity dispersion
and is expected to increase H$_0$ by 30--40\% (Wucknitz 2002). For
completeness, we note that a total mass profile with $\gamma'=2$ can
be excluded at the 99.9\%(99\%) CL for both the
isotropic(non-isotropic) JF and HQ models.

If we now include the constraints on the stellar (cluster)
mass-to-light ratio (\S 2.1; see TK02 or KT02 for details) -- which
gives the stellar mass fraction inside the Einstein radius independent
of H$_0$ -- we find dark-matter slopes of $\gamma$=$2.35\pm0.25$ (68\%
CL) and $\gamma$=$2.60\pm0.20$ for the isotropic JF (shown as an
example in Fig.~2) and HQ profiles, respectively. Mildly anisotropic
JF models with $r_i={\rm R}_e$ lead to $\gamma=2.20^{+0.35}_{-0.60}$,
whereas similar HQ models give $\gamma=2.50^{+0.3}_{-0.2}$. The
stellar mass fraction for a JF profile is $f_*=0.70^{+0.20}_{-0.25}$
(68\% CL; compare, e.g., the results in Bertola et al.\ 1993 for local
E/S0) inside the Einstein radius (0.05 lower for HQ profile). This
fraction lowers by 0.1 when we use the field M/L evolution (\S
2.1). 

For the HQ profile with its relatively shallow inner luminosity
density profile, the dark-matter halo density dominates at lens-galaxy
radii $\la$0.5~kpc (in fact, this happens even for HQ profiles in an
$r^{-2}$ total mass profile). For this reason we regard this a less
likely mass model, although we will continue to use it to illustrate
the effect of a range of inner luminosity-density profiles on the
determination of H$_0$.  A small core radius of $\sim 0.01$~kpc for
the steeper JF models is sufficient to avoid this problem and has only
negligible effect on the stellar dispersion. We also find that
$r^{-2}$ total mass profiles with neither a constant $\beta(r)$
(negative or positive, i.e. tangentially or radially anisotropic) nor
an Osipkov-Merritt model at the limit of radial instability
(e.g. Nipoti, Londrillo \& Ciotti 2002) can reproduce the observed
stellar velocity dispersion.

\vspace{-0.3cm}
\section{The Hubble Constant}
\label{sec:H0}

Having the constraints on $f_*$ and $\gamma$ in hand -- we can tightly
constrain the lens models and determine H$_0$ (\S 3), thereby
including in its error budget the uncertainties on the mass profile of
the lens.  We use the lens-code from Keeton (2001) to model the system
with the data described above.  As a test, we are able to recover the
results from I98 within the errors with both their astrometry and
ours. The stellar mass fraction and the slope of the dark-matter halo
inside the Einstein radius are set equal to those determined in the
previous paragraph. This leaves as free parameters: the ellipticity,
the position angle, and mass inside the Einstein Radius of the
dark-matter halo; the source position; the position and mass of the
nearby group.  For each mass model, we find the set of free parameters
that minimises the $\chi^2$. From the resulting minimum--$\chi^2$ lens
model, we determine H$_0$ by comparing the model time-delay between
images B and C with the observed time delay (\S 3). We stress that all
constraints on the lens model are independent of H$_0$.

\begin{table}
\label{tab:dH0}
\caption{Values of H$_0$ from PG1115+080. For each combination of mass
models for the lens galaxy and the nearby group we list the value of
H$_0$ in \hub\, along with its random uncertainty (JF=Jaffe,
HQ=Hernquist, ISO=isotropic, OM=Osipkov-Merrit anisotropy with
$r_i={\rm R}_e$, SIS=singular isothermal sphere, NFW=Navarro, Frenk \&
White).  
All models have $\chi^2$ in the range 2.8--2.9 (without the ring constraints).}
\begin{tabular}{lrrrr}
  \hline
      & JF-ISO & HQ-ISO & JF-OM & HQ-OM \\
  \hline
group SIS & 57$^{+12}_{-7}$ & 61$^{+12}_{-7}$ &  56$^{+11}_{-9}$ &  60$^{+12}_{-7}$\\
group NFW & 59$^{+12}_{-7}$ & 62$^{+14}_{-9}$ &  57$^{+12}_{-9}$ &  62$^{+12}_{-8}$\\
 \hline
\end{tabular}
\end{table}

Additional constraints can be obtained by using the shape of the
Einstein ring (Kochanek et al.\ 2001), as determined by tracing its
peak surface brightness in 47 independent points. We do not use the
brightness maxima or minima of the ring, which are difficult to
measure accurately. We assume an error of $0\farcs02$ on the ring
peak-brightness positions, chosen to yield a minimum-$\chi^2$
approximately equal to the number of degrees of freedom for the
$r^{-2}$ lens model. In this way, we find minimum--$\chi^2$ values of
56, 55 and 55 for models where the lens galaxy is modeled as a single
singular isothermal ellipsoid (I98), a single Pseudo-JF (i.e. constant
M/L) profile with half-light radius equal to ${\rm R}_e$
(i.e. $a\equiv(4/3)\,{\rm R}_e=1\farcs13$; see Keeton 2001), and our
best HQ model embedded in a steep dark-matter halo (see \S 2.2),
respectively (a singular isothermal group is assumed).  Remarkably,
the $\chi^2$ values are indistinguishable and the values of H$_0$ in
all three cases are found to be similar to those without the ring as
an additional constraint. We therefore conclude that the Einstein ring
does not allow us to distinguish between an $r^{-2}$ mass model, a
constant M/L pseudo-JF model with break radius $a<2\farcs0$, or our
best dynamical mass model, nor does it improve significantly the
accuracy on H$_0$ (see also Saha \& Williams 2001).  Finally, we note
that all our models give $r_{\rm ABC}=\Delta t_{\rm AC}/\Delta t_{\rm
BA}\approx 1.3$, consistent with Barkana (1997).

Tab.~3 lists the values of H$_0$ obtained for a variety of mass models
(HQ to JF luminous mass profiles, isothermal to NFW group mass
profiles, isotropic to non-isotropic stellar mass distributions), all
consistent with the available observations (see \S 2.3).  This
relatively large set of models leads to a tight range of
H$_0$=56--62~\hub. 

\begin{figure}
\begin{center}
\leavevmode
\hbox{%
\epsfxsize=\hsize
\epsffile{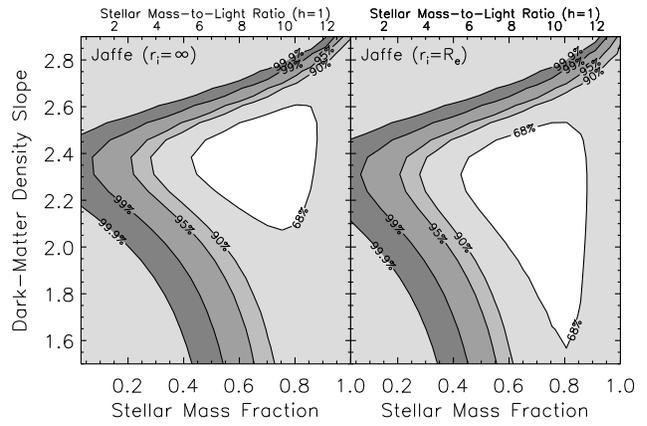}}
\end{center}
\caption{Likelihood contours of fraction of stellar mass within the
Einstein radius vs.\ dark matter slope ($\gamma$) for the isotropic (left panels)
and non-isotropic (right panels) JF model.  We note that both the slope of the 
dark-matter halo
($\gamma$) and the stellar mass fraction enclosed by the Einstein
radius are independent from H$_0$.}
\label{fig:mass}
\end{figure}

To compute the random errors we include the contributions from
uncertainty in the images and lens positions, in the time-delay, in
the image flux-ratios, in the slope of the dark-matter halo and in the
fraction of stellar mass inside the Einstein radius. The errors from
the lens model and dynamical model are determined separately, assuming
they are independent. In all cases the errors on H$_0$ due to the lens
models significantly dominate those due to the dynamical models. In
particular, the uncertainties on the galaxy position and time-delay
dominate the error budget on H$_0$. We combine the random errors from
the lensing and dynamical models and list the 68\% CL limits in
Tab.~3. Because we re-optimize $\chi^2$ until $\Delta\chi^2=1$, the 
error-ranges include the effects of any parameter degeneracies (i.e.
internal versus external shear) within
the context of our mass model.

We adopt, as best estimate of H$_0$ from PG1115, the average value from
the eight models considered in Table~\ref{tab:dH0}, for the random
uncertainty the average of the random uncertainties and for the
systematic uncertainty the semi-difference between the maximum and
minimum values, i.e. H$_0=59^{+12}_{-7}\,\pm3$~\hub\
(random/systematic).

\vspace{-0.3cm}
\section{Summary \& Discussion}

\label{sec:disc}

We have combined mass-to-light ratio constraints with lensing and
dynamical analyses -- using the stellar velocity dispersion measurement
from Tonry (1998) -- to constrain the radial mass profile of the lens
galaxy of PG1115+080 and break its degeneracy with the value of H$_0$, as
determined from the measured time delay. Our main results are:

\smallskip
{\bf (1)} A steep total mass profile with $\rho$$\propto$$r^{-\gamma'}$, 
where $\gamma'$=2.35$\pm$0.1$\pm$0.05, and an isotropic or mildly radial
velocity ellipsoid, successfully reproduces the stellar velocity
dispersion and the lensing constraints. A two-component mass model,
with a luminous component with stellar $M_*/L_B$ as determined from
the evolution of the FP, satisfies all the constraints only for a
relatively steep (compare TK02 and KT02) inner slope of the dark
matter halo and a considerable fraction of stellar mass within the
Einstein Radius, $f_*=0.67^{+0.2}_{-0.25}\pm0.03$.  This model
explains the offset of PG1115+080 from the FP in terms of structural
differences and does not require strong radial anisotropy of the
stellar velocity distribution.

In previous work, {\it ad hoc} distributions of stellar orbits have
been constructed inside a logarithmic potential also leading to a high
stellar velocity dispersion (Romanowsky \& Kochanek 1999). However, if
the enclosed mass is kept fixed (given by the lens model), these
orbits need to be extremely radial and instability issues may
arise. Recent numerical analyses show that if E/S0 galaxies are offset
from the FP through radial anisotropy by more than the observed
scatter in the FP, they become unstable and evolve back to the FP
(Nipoti et al.\ 2002). Tangential orbits always lead to a lower
central velocity dispersion. Our mass model explains the offset by
breaking the homology in the mass distribution, and is consistent with
the anisotropy structure expected in galaxy-formation scenarios, with
close to isotropic orbits in the central regions (e.g. van Albada
1982; see also the observations in Gerhard et al.\ 2001).

Our similar analyses of two other lenses (MG2016+112 and 0047--281;
TK02 and KT02) indicate that both lie on the FP of field E/S0 galaxies
{\sl and} not only have total mass profiles with slope $-$2 within 5\%,
but in the case of 0047--281, a spatially resolved velocity dispersion
profile allows us to rule out tangential anisotropy or significant
radial anisotropy in the center of the galaxy. Data for other lenses
with quality comparable to 0047--281 are being collected by the LSD
Survey, to provide more information on the mass distribution and
orbital structure of E/S0 (lens) galaxies.

\smallskip
{\bf (2)} Applying these dynamical constraints to the lens mass models
of PG1115+080, we find H$_0=59^{+12}_{-7}\pm3$~\hub\, including random
(68\% CL) and systematic uncertainties. To achieve higher accuracy it
will be necessary to further reduce the dominant sources of
uncertainty, i.e. errors on astrometry and time delay.  Note that this
value is significantly higher than the value of H$_0=44$$\pm$4~\hub\
found by I98, who assumed an $r^{-2}$ mass profile.

\smallskip
Comparing to other lensing measurements of H$_0$, our value is
consistent with those derived from B1608+656 (Fassnacht et al.\ 1999;
Koopmans \& Fassnacht 1999; Fassnacht et al.\ 2002) and B1600+434
(Koopmans et al.\ 2000; Kochanek 2002), both assuming $r^{-2}$ mass
profiles. However, in other cases values of H$_0\la50$\hub\ have been
found using $r^{-2}$ mass profiles as well (e.g. Kochanek 2002). We
therefore conclude that the $r^{-2}$ approximation might not always
appropriate for the determination of H$_0$ from time-delays, given the
intrinsic scatter in the total mass profile and fraction of dark matter
in the central 1--2\,R$_e$ of E/S0 galaxies (see e.g., Bertin et al.\
1994; Carollo et al.\ 1995; Gerhard et al.\ 2001, for ranges of mass
profiles of E/S0 galaxies), particularly if there is evidence for
offsets from the FP. We propose that the apparent differences between
H$_0$, inferred from different lens galaxies modelled with the same
mass profiles, is therefore the result of structural, {\sl not
kinematic}, non-homology of E/S0 galaxies, which might be related to
differences in their formation or interaction histories (e.g. field
versus group/cluster).  Hence it appears that additional constraints,
such as stellar kinematics, are essential to precisely determine the
mass profile -- hence H$_0$ -- for at least some lens-systems with
measured time-delays.

Finally, as compared to local determinations of H$_0$, our value is
consistent with H$_0=59\pm6$ \hub\ from Saha et al.\ (2001; and
references therein), but only marginally consistent with the final
value, H$_0=72\pm8$~\hub, from the HST Key-project (Freedman et al.\
2001).

\vspace{-0.5cm}
\section{Acknowledgments}
We thank E.~Agol, G.~Bertin, L.~Ciotti, R.~Ellis, and O.~Wucknitz for
their comments on this manuscript. Based on observations collected
with the NASA/ESA HST, obtained at STScI, which is operated by AURA,
under NASA contract NAS5-26555.



\begin{thebibliography}{}

\bibitem[Barkana 1997]{B97} Barkana R., 1997, ApJ, 489, 21

\bibitem[\pc{Bender et al.\ }1998]{B98} Bender R., Saglia R.~P.,
Ziegler B., Belloni P., Greggio L., Hopp U., Bruzual G., 1998, ApJ,
493, 529

\bibitem[Bertin et al. 1994]{B94} Bertin G. et al.\ 1994, A\&A, 292, 381

\bibitem[\pc{Bertola et al.\ } 1993]{B93} Bertola, F., Pizzella, A.,
Persic, M., Salucci, P., 1993, ApJ, 416, L45

\bibitem[Bullock et al.\ 2001]{B01} Bullock J.~S., Kolatt T.~S.,
Sigad Y., Somerville R.~S., Kravtsov A.~V., Klypin A.~A., Primack
J.~R., Dekel A., 2001, MNRAS, 321, 598

\bibitem[Carollo et al.\ 1995]{C95} Carollo C.~M., de Zeeuw P.~T., van der Marel R.~P., Danziger I.~J., Qian E.~E., 1995, ApJ, 441, L25

\bibitem[Courbin et al.\ 1997]{C97} Courbin F., Magain P., Keeton
C.~R., Kochanek C.~S., Vanderriest C., Jaunsen A.~O., Hjorth J.,
1997, A\&A, 324, L1

\bibitem[\pc{Djorgovski \& Davis } 1987]{FP1} Djorgovski S., Davis M.,
1987, ApJ, 313, 59

\bibitem[\pc{Dressler et al.\ } 1987]{FP2} Dressler A., Lynden-Bell
D., Burstein D., Davies R.~L., Faber S.~M., Terlevich R.~J., Wegner
G., 1987, ApJ, 313, 42

\bibitem[Fassnacht et al. 1999]{F99} Fassnacht C.~D., Pearson T.~J.,
Readhead A.~C.~S., Browne I.~W.~A, Koopmans L.~V.~E., Myers S.~T.,
Wilkinson P.~N., 1999, ApJ, 527, 498

\bibitem[Fassnacht et al. 2002]{} Fassnacht C.~D., Xanthopoulos E., 
Koopmans L.~V.~E., Rusin, D., 2002, ApJ, in press, astro-ph/0208420

\bibitem[Freedman et al. 2001]{WF01} Freedman W.~L., et al.\, 2001, ApJ, 553, 47

\bibitem[Fruchter \& Hook 2002]{FH02} Fruchter A.~S., Hook R.~N. 2002, PASP, 114, 144

\bibitem[Gerhard et al. 2001]{G01} Gerhard O., Kronawitter A.,
Saglia R.~P., Bender R., 2001, AJ, 121, 1936

\bibitem[Hernquist 1990]{H90} Hernquist L., 1990, ApJ, 356, 359

\bibitem[Impey et al. 1998]{I98} Impey C.~D., Falco E.~E., Kochanek
C.~S., Lehar J., McLeod B.~A., Rix H.-W., Peng C.~Y., Keeton
C.~R., 1998, ApJ, 509, 551 [I98]

\bibitem[Jaffe(1983)]{1983MNRAS.202..995J} Jaffe W.\ 1983, MNRAS, 202,
995

\bibitem[Keeton 2001]{Ke} Keeton C.~R., 2001, ApJ, submitted

\bibitem[Keeton \& Kochanek 1997]{KK} Keeton C.~R., Kochanek C.~S., 1997, ApJ, 487, 42 

\bibitem[\pc{Kelson et al.\ }2000b]{K2000b} Kelson D.~D., Illingworth
G.~D., van Dokkum P.~G., Franx M., 2000b, ApJ, 531, 184

\bibitem[Kochanek et al.~2000]{2000ApJ...543..131K} Kochanek C.~S., et al., 2000, ApJ,  543, 131

\bibitem[Kochanek, Keeton \& McLeod 2001]{KKM} Kochanek C.~S., Keeton
C.~R., McLeod B.~A., 2001, ApJ, 547, 50

\bibitem[Kochanek 2002]{Ko02} Kochanek C.~S.,  2002, ApJ, submitted

\bibitem[Koopmans 2001]{Ko01} Koopmans L.~V.~E., 2001, PASA, 18, 179

\bibitem[Koopmans \& Fassnacht (1999)]{KF99} Koopmans L.~V.~E.,
Fassnacht C.~D., 1999, ApJ, 527, 513

\bibitem[Koopmans \& Treu (2002)]{KT02} Koopmans L.~V.~E., Treu,
T. 2002, ApJ, submitted, astro-ph/0205281 [KT02]

\bibitem[Krist \& Hook (2001)]{KH01} Krist J., Hook R.\ 2001, ``The Tiny Tim user's guide''

\bibitem[Kronawitter 2000]{K00} Kronawitter, A., Saglia, R.~P., Gerhard, O., \& Bender, R. 2000, A\&AS, 144, 53

\bibitem[Merritt 1985]{M85a} Merritt D., 1985a, AJ, 90, 1027

\bibitem[Merritt 1985]{M85b} Merritt D., 1985b, MNRAS, 214, 25

\bibitem[Navarro et al.\ 1997]{NFW} Navarro J., Frenk C.~S., White S.~D.~M, 1997, ApJ, 490, 493 [NFW]

\bibitem[Nipoti, Londrillo \& Ciotti 2002]{NLC02} Nipoti C.,
Londrillo P., Ciotti L. MNRAS, 332, 901

\bibitem[Osipkov 1979]{O79} Osipkov L.~.P., 1979, Pis'ma Astron. Zh., 5, 77

\bibitem[Refsdal 1964]{R64} Refsdal S., 1964, MNRAS, 128, 295

\bibitem[Romanowsky \& Kochanek 1999]{RK99} Romanowsky A.~J., Kochanek C.~S., 1999, ApJ, 516, 18 

\bibitem[Saglia et al. 1992]{SBS} Saglia, R.~P., Bertin, G. \& Stiavelli, M. 1992, apj, 384, 433

\bibitem[Saha \& Williams 1997]{SW97} Saha P., Williams L.~L.~R., 1997, MNRAS, 292, 148

\bibitem[Saha \& Williams 2001]{SW01} Saha P., Williams L.~L.~R., 2001, AJ, 122, 585

\bibitem[Saha et al. 2001]{Sa01} Saha A., Sandage A., Tammann
G.~A., Dolphin A.~E., Christensen J., Panagia N., Macchetto
F.~D. 2001, ApJ, 562, 314

\bibitem[Schechter et al. 1997]{S97} Schechter P.~L. et al., 1997, ApJ, 475, L85 

\bibitem[\pc{Schlegel et al.\ }1998]{EXMAPS} Schlegel D.~J., Finkbeiner D.~P., Davis M., 1998, ApJ, 500, 525

\bibitem[Tonry 1998]{T98} Tonry J.~L., 1998, AJ, 115, 1 

\bibitem[Treu \& Koopmans (2002)]{TK02} Treu T., Koopmans
L.~V.~E., 2002, ApJ, 575, 87 [TK02]

\bibitem[Treu et al.\ (2001a)]{T01a} Treu, T., Stiavelli, M., Bertin G., Casertano, C., \& M{\o}ller, P. 2001a, MNRAS, 326, 237

\bibitem[\pc{Treu et al.\ }2002]{T02} Treu T., Stiavelli M., Casertano
S., M\o ller P., Bertin G., 2002, ApJ, 564, L12

\bibitem[\pc{Treu et al.\ }2001]{T01b} Treu T., Stiavelli M., M\o ller
P., Casertano S., Bertin G., 2001b, MNRAS, 326, 221 

\bibitem[\pc{van Albada }1982]{vA82} van Albada T.~S., 1982, MNRAS, 201, 939

\bibitem[\protect\citename{van Dokkum et al.\ }1998]{DFKI98} van
Dokkum P., Franx M., Kelson D.~D., Illingworth G.~D., 1998, ApJ, 504,
L17

\bibitem[Weynman et al. 1980]{W80} Weynman R.~J. et al.\ 1980,
Nature, 285, 61

\bibitem[Williams \& Saha 2000]{WS00} Williams L.~L.~R., Saha P.,
2000, AJ, 119, 439

\bibitem[Wucknitz 2002]{2002MNRAS.332..951W} Wucknitz O., 2002, MNRAS,  
332, 951 

\bibitem[Zaho \& Pronk 2001]{ZP01} Zaho H., Pronk D., 2001, MNRAS, 320, 401

\end{thebibliography}
\end{document}